\documentclass[aps,prb,reprint]{revtex4-2}

\usepackage{graphicx}
\usepackage{amsmath}   
\usepackage{amssymb}   

\begin{document}

\title{Strain-Tunable Harmonic Responses in Valley-Polarized Bilayer Graphene}

\author{Narjes Kheirabadi, Aliasghar Shokri }
\affiliation{Department of Theoretical and Nanophysics, Faculty of Physics, Alzahra University, Tehran, Iran}

\begin{abstract}
We theoretically investigate the linear and second-order nonlinear optical responses of valley-polarized bilayer graphene under uniaxial strain. Employing a low-energy effective Hamiltonian that incorporates trigonal warping and strain-induced anisotropy, we calculate the optical susceptibilities within the quantum kinetic formalism. We show that, while the second-order response vanishes in valley-balanced bilayer graphene owing to the cancellation of contributions from opposite valleys, a finite valley polarization lifts this cancellation and enables a net second-harmonic generation (SHG) signal. Uniaxial strain substantially modifies the nonlinear response by distorting the low-energy electronic structure and altering the pseudospin texture, producing a highly anisotropic SHG spectrum. Pronounced resonant enhancements occur at photon energies $\hbar\omega \approx E_f$ and $\hbar\omega \approx 2E_f$, associated with two-photon and one-photon interband resonances, respectively. Remarkably, changing the sign of the strain parameter reverses the direction of the induced second-harmonic current, providing a mechanically controlled switching mechanism for nonlinear optical transport. These results establish strain engineering as an effective route for manipulating valley-dependent nonlinear optical phenomena in bilayer graphene and suggest new opportunities for tunable mid-infrared photonic and valley-optoelectronic applications.
\end{abstract}

%
\maketitle
\pagebreak

\section{Introduction}
Bilayer graphene provides a highly versatile platform for investigating symmetry-controlled electronic and optical phenomena in two-dimensional quantum materials \cite{McCann2013BLGReview,Kheirabadi2018Cyclotron,
Kheirabadi2016MagneticRatchet,Battilomo2019, Kheirabadi2022QuantumNonlinear,
Kagan2025BilayerGraphene,Li2026TwistedBilayerSystems}. In contrast to monolayer graphene, BLG hosts massive chiral quasiparticles and possesses a low-energy band structure that can be continuously tuned by external perturbations such as electric fields, strain, pressure, and stacking modifications \cite{McCann2013BLGReview, MuchaKruczynski2010, Rozhkov2016}. Owing to the interplay between interlayer coupling and trigonal warping, the electronic structure of BLG exhibits a rich variety of phenomena, including Lifshitz transitions, anisotropic transport, and tunable optical responses \cite{Mucha2011,McCann2013BLGReview, McCann2012ElectronicProperties}. These properties have made bilayer graphene an attractive system for exploring nonlinear optical effects and their manipulation through symmetry engineering.

Recent advances in nonlinear optics of quantum materials have generated significant interest in second-order optical processes, including second-harmonic generation (SHG), nonlinear photocurrents, and nonlinear Hall effects \cite{Sodemann2015,Shen2019}. In centrosymmetric crystals, electric-dipole second-order responses vanish because of inversion symmetry. Consequently, pristine bilayer graphene exhibits negligible SHG under ordinary conditions. Finite second-order optical responses can emerge only when inversion symmetry or other relevant symmetries are broken, for example through perpendicular electric fields, substrate-induced asymmetry, twist engineering, strain, or valley-selective population imbalance \cite{Brun2015, Yao2008, Hunt2013, Ju2015}. Understanding how such symmetry-breaking mechanisms influence nonlinear optical phenomena remains a central topic in graphene-based optoelectronics.

Several routes for activating SHG in graphene systems have been explored. In electrically biased bilayer graphene, an interlayer potential difference breaks inversion symmetry and produces a strong, tunable SHG response in the mid-infrared regime \cite{Brun2015}. Twisted bilayer graphene can exhibit enhanced SHG arising from the reduced symmetry of moiré superlattices and the presence of van Hove singularities \cite{Yang2020}. More recently, strain engineering has been demonstrated as an effective method for modifying optical selection rules and enhancing nonlinear optical responses through lattice distortions and pseudospin redistribution \cite{Lu2023StrongSHG}. it has been also shown that non-uniform strain breaks graphene's inversion symmetry through sublattice polarization, enabling a remarkably strong and tunable second-harmonic optical response in the visible range  \cite{Lu2023StrongSHG}. These studies collectively demonstrate that symmetry control provides a powerful mechanism for manipulating nonlinear optical phenomena in graphene-based materials.

Among the available tuning parameters, uniaxial strain is particularly appealing because it directly modifies the low-energy electronic structure without introducing additional material complexity \citep{Battilomo2019, Mucha2011}. In bilayer graphene, strain alters intralayer hopping amplitudes, distorts trigonal-warped energy contours, shifts the positions of Dirac points, and changes the topology of the Fermi surface near the Lifshitz transition \cite{Pereira2009, Guinea2010, Mucha2011}. Such modifications can strongly influence optical matrix elements and resonant interband transitions. Despite extensive studies of strain effects on the electronic structure and transport properties of bilayer graphene, their consequences for nonlinear optical responses remain insufficiently understood, particularly in the presence of valley-selective carrier populations.

\begin{figure}[t!]
   \centering   
   \includegraphics[scale=0.2]{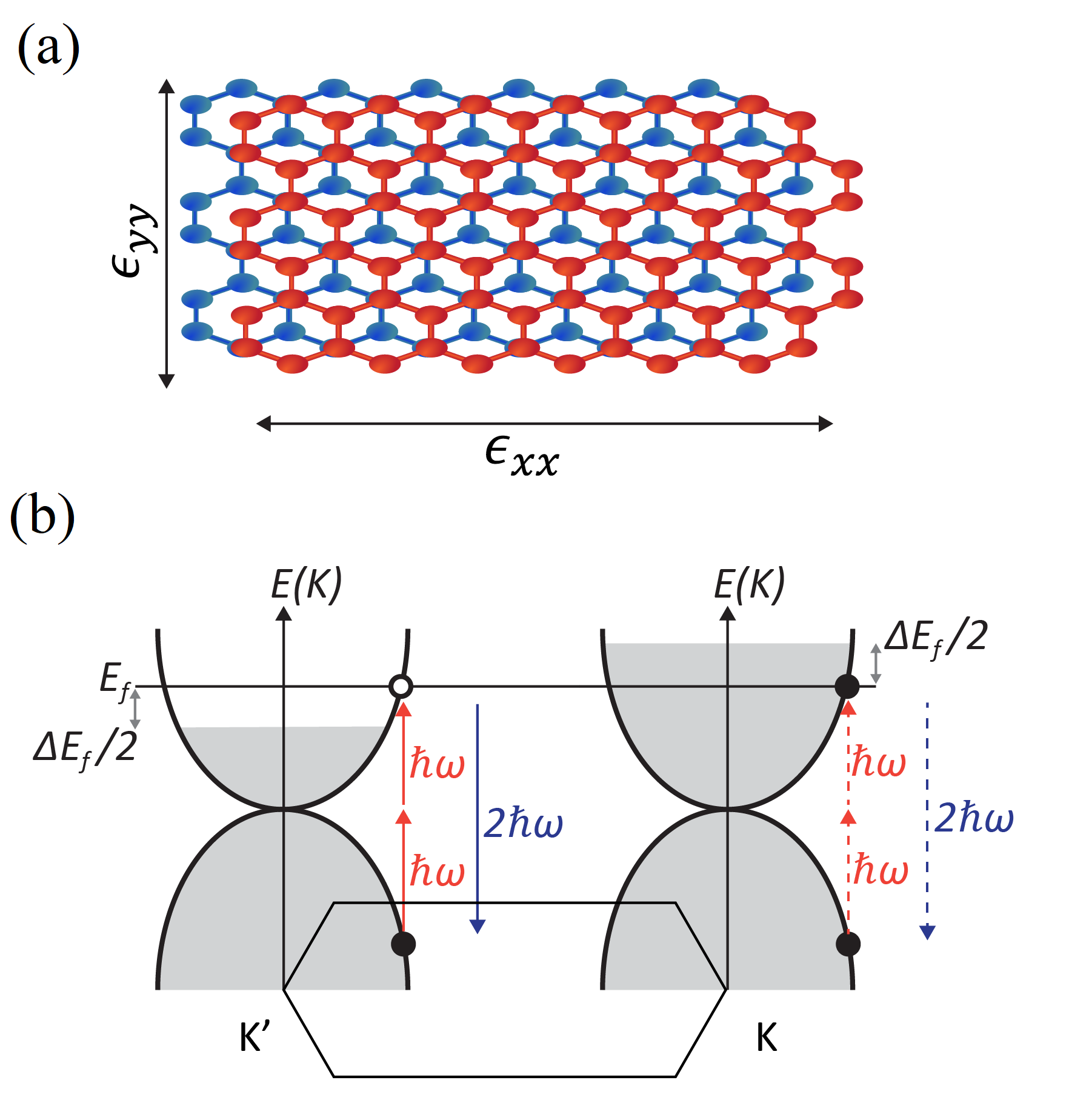}
   \caption{Strained bilayer graphene and valley polarization. (a) Schematic illustration of uniaxially strained bilayer graphene, where $\epsilon_{xx}$ and $\epsilon_{yy}$ denote the strain components along the respective axes. (b) Energy band structure of the $K$ and $K'$ valleys featuring valley polarization. The shifting Fermi levels ($E_f \pm \Delta E_f/2$) and asymmetrical optical transitions ($\hbar\omega$ and $2\hbar\omega$) illustrate the intervalley population imbalance.}
    \label{0}
\end{figure}

Valley polarization introduces an additional symmetry-breaking mechanism that can fundamentally modify nonlinear optical processes \cite{tara}. In valley-balanced systems, contributions from the two inequivalent valleys are constrained by time-reversal symmetry and often cancel in the total second-order response. A finite valley polarization lifts this cancellation and permits a net nonlinear optical current \cite{tara}. The resulting response is expected to be especially sensitive to trigonal warping and strain-induced anisotropy, both of which modify the valley-resolved pseudospin texture and optical transition amplitudes. Consequently, the combined action of valley polarization and strain offers a promising route toward controlling nonlinear optical phenomena through both electronic and mechanical degrees of freedom (Fig.~\ref{0}).
Motivated by these considerations, we investigate the linear and second-order optical responses of valley-polarized bilayer graphene subjected to uniaxial strain. Using an effective low-energy two-band Hamiltonian that incorporates trigonal warping and strain-induced anisotropy, we solve the quantum kinetic equation within the density matrix formalism and evaluate the valley-resolved linear and nonlinear optical conductivities. We demonstrate that, while the second-order response vanishes in valley-balanced bilayer graphene because of the cancellation between opposite valleys, a finite valley polarization activates a measurable SHG signal. We further show that uniaxial strain provides an efficient mechanism for tuning both the magnitude and anisotropy of the nonlinear response by reshaping the low-energy band structure and pseudospin texture. Remarkably, the sign of the strain-induced anisotropy can reverse the direction of the nonlinear current, providing a mechanically controlled switching mechanism for SHG. Our results establish strain engineering as an effective strategy for manipulating valley-dependent linear and nonlinear optical responses in bilayer graphene and highlight the potential of mechanically tunable SHG for mid-infrared photonic and optoelectronic applications.

\section{Hamiltonian and Theoretical Framework}
The low-energy electronic structure of strained bilayer graphene is described by the effective two-band Hamiltonian~\cite{Mucha2011, Battilomo2019}
\begin{equation}\label{H}
\begin{split}
H = & \left[ -\frac{1}{2m}(k_x^2-k_y^2) + \nu v_3 k_x + w \right]\sigma_x \\
    & - \left[ \frac{1}{m}k_xk_y + \nu v_3 k_y \right]\sigma_y + \frac{\Delta}{2}\sigma_z ,
\end{split}
\end{equation}
where $\sigma_i$ are the Pauli matrices acting in the low-energy sublattice space, and $\nu=\pm1$ labels the $K$ and $K'$ valleys, respectively. 
Here, $m=\gamma_1/2v_0^2$ is the effective mass, $v_0=\sqrt{3}a\gamma_0/2\hbar$ denotes the Fermi velocity, and $v_3=\sqrt{3}a\gamma_3/2\hbar$ accounts for trigonal warping, with $a = 2.46\,\text{\AA}$ being the lattice constant of pristine bilayer graphene. The parameters $\gamma_0$, $\gamma_1$, and $\gamma_3$ denote the intralayer nearest-neighbor, vertical interlayer, and skew interlayer hopping amplitudes, with representative values of $3.16\,\text{eV}$, $0.39\,\text{eV}$, and $0.32\,\text{eV}$, respectively. The parameter $\Delta$ represents the inversion-symmetry-breaking gap induced, for instance, by an perpendicular interlayer electric field. 

Uniaxial strain enters the effective Hamiltonian through the symmetry-breaking term $w\sigma_x$, where
\begin{equation}
w = \frac{3}{4}\gamma_3(\epsilon_{xx}-\epsilon_{yy})(\beta_3-\beta_0).
\end{equation}
Here, $\epsilon_{ij}$ is the strain tensor, and the difference $\epsilon_{xx}-\epsilon_{yy}$ characterizes the anisotropic lattice deformation associated with the uniaxial strain. 
Physically, a $1\%$ uniaxial strain corresponds to a relative change of approximately $0.01$ in the lattice spacing along a primary crystallographic direction (e.g., $\epsilon_{xx}=0.01$). Due to the Poisson effect, the transverse direction undergoes a smaller compression, $\epsilon_{yy} \approx -\sigma_{\text{P}}\epsilon_{xx}$, where $\sigma_{\text{P}} \approx 0.16$--$0.18$ is the Poisson's ratio of AB-stacked bilayer graphene~\cite{Wei2009ElasticGraphene, Blakslee1970GraphiteElastic, Lee2008MeasurementGraphene}. Consequently, the net anisotropic strain scaling the effective Hamiltonian reduces to $\epsilon_{xx}-\epsilon_{yy} \approx (1+\sigma_{\text{P}})\epsilon_{xx}$. 

The dimensionless quantities $\beta_i = -\partial\ln\gamma_i / \partial\ln a$ are the Grüneisen parameters describing the strain dependence of the respective hopping amplitudes, with typical values spanning $\beta_0 \sim 2$--$3$ and $\beta_3 \sim 4$--$6$~\cite{CastroNeto2009ReviewGraphene, Guinea2010}. The parameter $w$ therefore serves as a direct metric for the strain-induced anisotropy of the low-energy electronic structure, governing the distortion of the trigonally warped Fermi surface near the Lifshitz transition. The sign of $w$ determines the orientation of this anisotropy: $w > 0$ corresponds to $\epsilon_{xx} > \epsilon_{yy}$ (predominant tensile deformation along the $x$ axis), whereas $w < 0$ implies $\epsilon_{xx} < \epsilon_{yy}$, yielding an inverted orientation of the Fermi surface deformation. The strain values considered in this work correspond to realistic uniaxial deformations on the order of $0.1\%$--$2\%$, which are routinely achievable in graphene via substrate engineering, flexible polymer supports, and microelectromechanical systems (MEMS) techniques~\cite{Guinea2010, naumis2017strain, pereira2009strain}. 

We assume an in-plane, linearly polarized ($p$-polarized) incident radiation field given by $\mathbf{E}(t) = 2 \mathbf{E}(\omega) \cos(\omega t)$. The application of this in-plane field induces a time-dependent polarization $\mathbf{p}(t)$ within the material. This polarization drives a macroscopic current density defined by $\mathbf{j}(t) = \partial \mathbf{p}(t)/\partial t$. Consequently, the current density can be expanded in terms of its frequency components as
\begin{equation}\label{ro}
\begin{split}
\mathbf{j}(t) = & \left[ -i \omega \mathbf{p}(\omega) e^{-i \omega t} + \text{c.c.} \right] \\
& + \left[ -2i \omega \mathbf{p}(2\omega) e^{-2i \omega t} + \text{c.c.} \right] + \dots,
\end{split}
\end{equation}
where $\text{c.c.}$ denotes the complex conjugate.

To investigate the mid-infrared optical response of strained bilayer graphene, we employ the quantum kinetic equation for the density matrix $\rho$,
\begin{equation}\label{qke}
\frac{\partial\rho}{\partial t} = -\frac{i}{\hbar} [H + V, \rho] + \mathrm{St}\rho,
\end{equation}
where $H$ is the unperturbed Hamiltonian and $V$ is the electron-photon interaction Hamiltonian evaluated in the velocity gauge (minimal coupling). The interaction term is given explicitly by
\begin{equation}\label{v}
V = -\frac{e}{c} \mathbf{v} \cdot \mathbf{A} + \frac{e^{2}}{2c^{2}} \sum_{\alpha\beta} \frac{\partial v_{\alpha}}{\partial k_{\beta}} A_{\alpha} A_{\beta},
\end{equation}
where $\mathbf{A}$ is the vector potential of the electromagnetic field. The time-dependent vector potential is expressed as $\mathbf{A}(t)=\mathbf{A}(\omega) \exp(-i \omega t) + \mathbf{A}^*(\omega) \exp(i \omega t)$, which, for linearly polarized light, simplifies via the relation $\mathbf{A} (\omega) = -i(c/\omega) \mathbf{E}(\omega)$. The collision integral, $\mathrm{St} \rho$, accounts for the relaxation effects, where the interband contributions to the density matrix decay at a characteristic rate of $\gamma/ \hbar$. In all numerical calculations, we take $ = 10$ meV.  

The time-dependent density matrix $\rho(t)$ solving the quantum kinetic equation can be expanded perturbatively as
\begin{equation}\label{jro}
\rho (t) = \rho^{(0)} + [\rho^{(1)} e^{-i\omega t} + \mathrm{c.c.}] + [\rho^{(2)} e^{-2i\omega t} + \mathrm{c.c.}] + \dots
\end{equation}
Taking $-e$ ($e>0$) as the electron charge, the macroscopic current density $\mathbf{j}(t)$ is given by
\begin{equation}\label{jt}
\mathbf{j}(t) = - e \sum_{\mathbf{k},\nu} \mathrm{Tr} [\rho(t) \mathbf{v}],
\end{equation}
where the velocity operators are defined as
\begin{equation}\label{v_op}
\hat{v}_\alpha = \frac{1}{\hbar}\frac{\partial H(\mathbf{k})}{\partial k_\alpha}, \qquad \alpha = x,y.
\end{equation}

By matching the current expansion with the definition of the time-dependent polarization ($\mathbf{j} = \partial \mathbf{p}/\partial t$), the frequency-dependent polarization components at the fundamental ($\omega$) and second-harmonic ($2\omega$) frequencies are found to be
\begin{equation}\label{pomega}
\mathbf{p} (\omega)=\frac{e g_s}{i \omega } \sum_{\mathbf{k}, \nu} {\mathrm{Tr} [\rho^{(1)} \mathbf{v}]},
\end{equation}
\begin{equation}\label{p2omega}
\mathbf{p} (2\omega)=\frac{e g_s}{2 i \omega } \sum_{\mathbf{k}, \nu} {\mathrm{Tr} [ \rho^{(2)} \mathbf{v} ]},
\end{equation}
where $g_s$ is the spin degeneracy factor and $\nu$ labels the valley index.
For a two-dimensional material subjected to an in-plane driving electric field, the linear and second-harmonic polarization components are given by
\begin{align}
p_\alpha(\omega) &= \chi_{\alpha\beta}(\omega) E_\beta(\omega), \label{eq:linear_pol} \\
p_\alpha(2\omega) &= \chi_{\alpha\beta\gamma}(2\omega; \omega, \omega) E_\beta(\omega) E_\gamma(\omega), \label{eq:shg_pol}
\end{align}
where the Cartesian indices $\alpha, \beta, \gamma \in \{x, y\}$ denote the in-plane coordinates, and $\chi_{\alpha\beta}$ and $\chi_{\alpha\beta\gamma}$ represent the linear and second-order nonlinear optical susceptibility tensors, respectively. 

Substituting Eq.~\eqref{jro} into the quantum kinetic equation~\eqref{qke} yields the first-order interband coherence element
\begin{equation}\label{rho1}
\rho^{(1)}_{cv} = \frac{i e}{\omega} \frac{ (\mathbf{E}(\omega) \cdot \mathbf{v}_{cv}) (f_v - f_c) }{ \hbar\omega - E_{cv} + i\gamma },
\end{equation}
where $f_{c,v}$ are the equilibrium Fermi-Dirac distribution functions for the conduction and valence bands, respectively, and $E_{cv} = E_c - E_v$ is the interband transition energy. The relation $\rho^{(1)}_{vc}(\omega) = [\rho^{(1)}_{cv}(-\omega)]^*$ ensures that the time-domain density matrix $\rho(t)$ remains Hermitian. 

For the effective two-band Hamiltonian considered here, the first-order corrections to the diagonal elements of the density matrix vanish identically, i.e., $\rho_{cc}^{(1)} = \rho_{vv}^{(1)} = 0$. Notably, this result holds independently of strain, trigonal warping, or the presence of an inversion-symmetry-breaking gap. The linear optical response is therefore entirely governed by the interband coherences $\rho_{cv}^{(1)}$ and $\rho_{vc}^{(1)}$, meaning that only interband transitions contribute to the linear current density.

The linear optical susceptibility is a second-rank tensor~\cite{Boyd2020NonlinearOptics}. Time-reversal symmetry constraints dictate that the tensor is symmetric, i.e., $\chi_{ij}^{(1)} = \chi_{ji}^{(1)}$. Furthermore, under the $C_{2}(x)$ point-group symmetry operation, the linear susceptibility tensor for the strained bilayer graphene reduces to a purely diagonal form:
\begin{equation}
\chi^{(1)}(\omega) = \begin{pmatrix} \chi_{xx}(\omega) & 0 \\ 0 & \chi_{yy}(\omega) \end{pmatrix}.
\end{equation}
This structure indicates an anisotropic yet decoupled optical response along the principal axes. Consequently, an in-plane driving electric field polarized along the $x$ direction induces an optical polarization exclusively along the $x$ axis, and an identical restriction applies to fields polarized along the $y$ axis. Off-diagonal cross-coupling is thus strictly forbidden at the linear response level.

For strained bilayer graphene, where the two-fold rotation axis lies along the $x$ direction within the $xy$ plane of the monolayer, the second-order nonlinear polarization induced by an in-plane electric field is strictly confined to the 2D plane. The nonvanishing components of the second-harmonic polarization are given explicitly by
\begin{align}
p_x(2\omega) &= \chi_{xxx} E_x^2 + \chi_{xyy} E_y^2, \label{eq:px2w} \\
p_y(2\omega) &= 2\chi_{yxy} E_x E_y, \label{eq:py2w}
\end{align}
where the factor of $2$ accounts for the intrinsic permutation symmetry of the susceptibility tensor with respect to its last two indices (i.e., $\chi_{ijk} = \chi_{ikj}$ for $j \neq k$). 

This macroscopic polarization relation matches the microscopic description obtained from Eq.~\eqref{p2omega}. By solving the quantum kinetic equation to second order, the nonvanishing interband and intraband density matrix elements are found to be
\begin{equation}\label{rho2}
\begin{split}
\rho_{cv}^{(2)} = & \frac{2 (e / \omega)^2( \mathbf{E}(\omega) \cdot \mathbf{v}_{cv})(\mathbf{E}(\omega) \cdot \mathbf{v}_{cc})}{(2\hbar\omega - E_{cv} + i\gamma)(\hbar\omega + E_{cv} + i\gamma)}(f_c - f_v) \\
& +\frac{(e / \omega)^2 \sum_{\alpha \beta} \left( \frac{\partial v_{\alpha}}{\partial k_{\beta}} \right)_{cv} E_{\alpha}(\omega) E_{\beta}(\omega)}{2(2\hbar\omega - E_{cv} + i\gamma)}(f_c - f_v),
\end{split}
\end{equation}
and
\begin{equation}
\rho_{cc}^{(2)} = -\frac{(e/\omega)^2(\mathbf{E}(\omega) \cdot \mathbf{v}_{cv})(\mathbf{E}(\omega) \cdot \mathbf{v}_{vc})}{(\hbar\omega - E_{cv} + i\gamma)(\hbar\omega + E_{cv} + i\gamma)}(f_c - f_v),
\end{equation}
respectively.
Equations~\eqref{rho2} are derived by substituting the perturbative expansion of Eq.~\eqref{jro} into the quantum kinetic equation~\eqref{qke}~\cite{tara}. Here, electron-hole symmetry dictates that the diagonal velocity matrix elements satisfy $\mathbf{v}_{cc} = -\mathbf{v}_{vv}$. Furthermore, hermiticity and conservation of total particle number require that the second-order density matrix elements obey the relations $\rho_{vc}^{(2)}(\omega) = [\rho_{cv}^{(2)}(-\omega)]^*$ and $\rho_{cc}^{(2)} = -\rho_{vv}^{(2)}$, respectively. Utilizing these symmetry constraints, the second-order density matrix elements in Eq.~\eqref{p2omega} can be uniquely determined.

The linear and second-order nonlinear susceptibilities are calculated for each valley independently, with the total optical response obtained from the coherent summation over both valleys. For the low-temperature degenerate semiconductor regime considered here, the low-energy optical transitions are predominantly dictated by electronic states in the immediate vicinity of the Fermi surface. To isolate these dominant contributions, we restrict the momentum summation to a narrow energy window around the Fermi energy $E_f$, defined by $E_f < E(\mathbf{k}) \le E_f + \delta E$. We set $\delta E = 1\,\text{meV}$, and have verified that the numerical results are insensitive to moderate variations of this cutoff, thereby confirming the convergence of our calculations. 

At finite temperatures, interband transitions are naturally weighted by the full Fermi-Dirac occupation factor difference, $f_c - f_v$, which accounts for thermal broadening effects. All numerical evaluations are performed at a representative temperature of $T = 5\,\text{K}$. The optical responses are computed from Eqs.~\eqref{pomega} and \eqref{p2omega} by performing Brillouin zone integrations over a discrete $500 \times 500$ $\mathbf{k}$-point grid~\cite{Kheirabadi2022QuantumNonlinear}, using a frequency step size of $\Delta(\hbar\omega/E_f) = 0.05$. 

It is worth noting that in uniaxially strained bilayer graphene, omitting either valley polarization or the skew interlayer hopping parameter $\gamma_3$ leads to an exact cancellation of the nonlinear response between the two valleys. The physical origin of the nonvanishing SHG current lies in the simultaneous presence of a valley population imbalance and the trigonal warping factor $v_3$. 

To model this nonequilibrium valley-polarized state, we implement a valley-dependent Fermi energy, $E_f(\nu) = E_f + \nu \Delta E_f / 2$, where $\nu = \pm 1$ is the valley index and $\Delta E_f$ quantifies the chemical potential mismatch between the two valleys~\cite{tara} (see Fig.~\ref{0}). Experimentally, such a valley polarization can be dynamically generated via circularly polarized optical pumping, nonlocal valley-Hall transport, or proximity-induced symmetry breaking in van der Waals heterostructures~\cite{hendry2010coherent}. Within this phenomenological framework, $\Delta E_f$ serves as a direct metric for the valley population imbalance, thereby controlling the magnitude of both the net linear and second-order optical responses. 

Here, we set $E_f = 100\,\text{meV}$ as a representative baseline Fermi level~\cite{McCann2013BLGReview}. The choice of $\Delta E_f = 1\,\text{meV}$ represents a weak valley polarization regime. This allows us to investigate the sensitivity of the nonlinear optical response to experimentally realistic, subtle valley imbalances that do not significantly perturb the underlying electronic structure, while still generating a measurable nonlinear signal. Because the total SHG signal is directly proportional to the population difference between the valleys, larger values of $\Delta E_f$ are expected to further scale and enhance the magnitude of the nonlinear susceptibility. Consequently, the predicted strain-controlled SHG response
should be experimentally detectable using modern
mid-infrared nonlinear optical spectroscopy techniques.

The existence of a finite second-order optical response in our model can be understood directly from symmetry considerations. For unbiased bilayer graphene ($\Delta=0$), the lattice retains global inversion symmetry, causing the macroscopic electric-dipole second-order response to vanish when the carrier populations of the two valleys are balanced. Locally, however, the low-energy Hamiltonian of each valley contains a trigonal warping term ($v_3 \neq 0$), which introduces a threefold anisotropy in momentum space and yields a finite, valley-resolved nonlinear susceptibility. Time-reversal symmetry maps the two valleys onto each other and dictates that their second-order susceptibilities satisfy the constraint $\chi^{(2)}(K) = -\chi^{(2)}(K')$. Consequently, in the absence of a valley population imbalance, the nonlinear signals from the two valleys exactly cancel, resulting in a vanishing macroscopic SHG response. Introducing a finite valley polarization lifts this cancellation by assigning different statistical weights to the valley occupations, thereby generating a nonvanishing net second-order susceptibility. Thus, the SHG response investigated in this work originates from the cooperative effect of trigonal warping and valley polarization rather than from an interlayer inversion-symmetry-breaking gap.

In addition, a finite interlayer asymmetry gap $\Delta$ is known to reduce the overall magnitude of the nonlinear response by suppressing low-energy optical transitions near the Fermi surface. Consequently, in the present study, we focus on the inversion-symmetric regime by setting the interlayer asymmetry gap to zero, thereby omitting the explicit effects of the inversion-symmetry-breaking parameter $\Delta$ [see Eq.~\eqref{H}].

\section{Results and discussions}
As outlined in the preceding section, the structural effect of uniaxial strain on bilayer graphene is modeled by introducing an explicit gauge field term, $w \sigma_x$, into the low-energy effective Hamiltonian. In the quantum kinetic framework employed here, the induced $n\text{th}$-order optical current densities are obtained from the density matrix elements via
\begin{equation}
\mathbf{j}^{(n)} \propto \text{Tr}\big[ \rho^{(n)} \hat{\mathbf{v}} \big],
\end{equation}
where $n = \{1, 2\}$, $\rho^{(n)}$ denotes the corresponding $n\text{th}$-order perturbative density matrix describing the system's dynamic response to the external driving field, and the velocity operator $\hat{\mathbf{v}}$ inherits the characteristic pseudospin structure of the strained Hamiltonian. To sweep a representative range of experimentally accessible, strain-induced symmetry-breaking configurations, we vary the strain energy parameter across $w = \pm 5$, $\pm 10$, and $\pm 15\,\text{meV}$. These values correspond to low-to-moderate physical strain levels in the $0.1\%$--$2\%$ regime, which are typically achievable in realistic substrate-deformation or suspended-membrane setups~\cite{Guinea2010, McCann2013BLGReview}.

For small values of the strain parameter $w$ and trigonal warping velocity $v_3$, the low-energy dispersion relation for the conduction band can be approximated as
\begin{equation}
E(\mathbf{k}) \approx \frac{k^2}{2m} - \big[ \nu k v_3 \cos(3\phi) + w \cos(2\phi) \big],
\end{equation}
where $m$ is the effective mass, $\nu = \pm 1$ is the valley index, and $\phi = \arctan(k_y/k_x)$ denotes the azimuthal angle in the momentum plane. Both the $v_3$-dependent trigonal warping term and the $w$-dependent strain term introduce a pronounced directional anisotropy, shifting the local conduction and valence band profiles as a function of the in-plane momentum direction. 

\begin{figure*}[t!]
   \centering
   \includegraphics[width=1\textwidth]{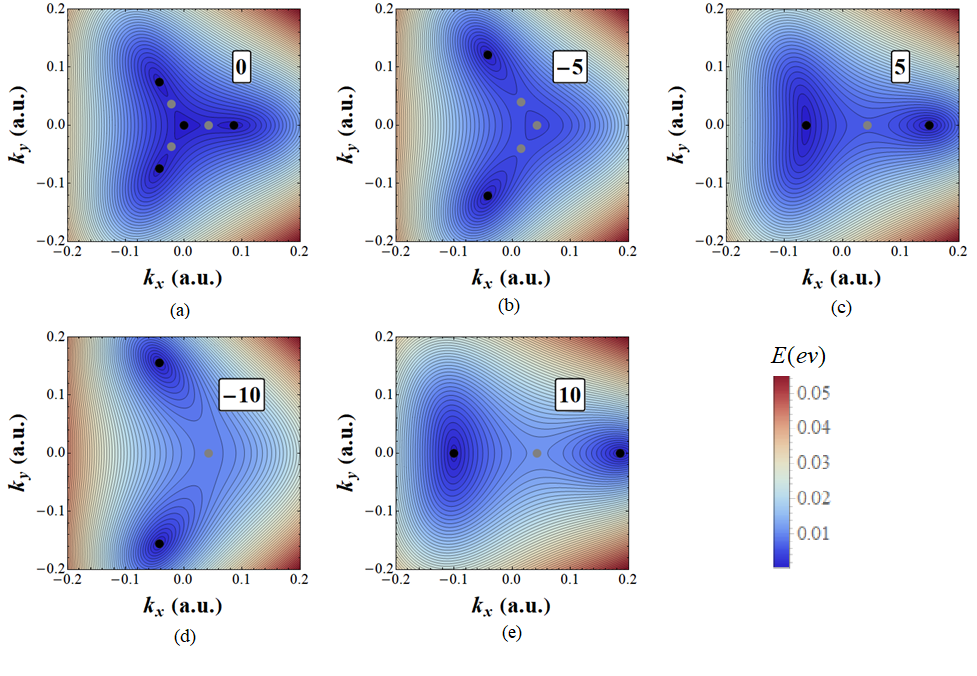}
   \caption{Constant-energy contours of the conduction band in the $K$ valley of uniaxially strained bilayer graphene. Panels (a)--(e) depict the evolution of the Fermi surface topology for the specified values of the strain parameter $w$ (in units of $\text{meV}$). The stable Dirac points and the saddle points are designated by black and gray dots, respectively.}
   \label{1}
\end{figure*}

To visualize this anisotropic deformation, we plot the constant-energy contours for four representative strain configurations in Fig.~\ref{1}. While the general influence of a strain-induced gauge field $w$ on the Lifshitz transition in bilayer graphene has been explored previously~\cite{Mucha2011}, Fig.~\ref{1} illustrates the evolution of the Fermi-surface topology and the migration of the Dirac points as a function of the strain parameter $w$.
As demonstrated in Fig.~\ref{1}, the strain-displaced Dirac cones induce a pronounced anisotropic deformation of the local Fermi surface that depends sensitively on the sign of the strain parameter $w$. For $w > 0$, the constant-energy contours and the associated Dirac points are elongated predominantly along the $k_x$ axis. Conversely, for $w < 0$, the structural deformation and point migration occur primarily along the $k_y$ direction.

For strained bilayer graphene, the parameter $w$ enters the Hamiltonian through the $\sigma_x$ term and therefore controls the directional character of the low-energy eigenstates. When $w>0$, the $\sigma_x$ contribution to the eigenstates is enhanced and the Dirac points are displaced along the $k_x$ direction. In contrast, for $w<0$, the $\sigma_x$ term is reduced as the magnitude of $w$ increases. 

\subsection{First order response}
\begin{figure}[t!]
   \centering   
   \includegraphics[width=0.48\textwidth]{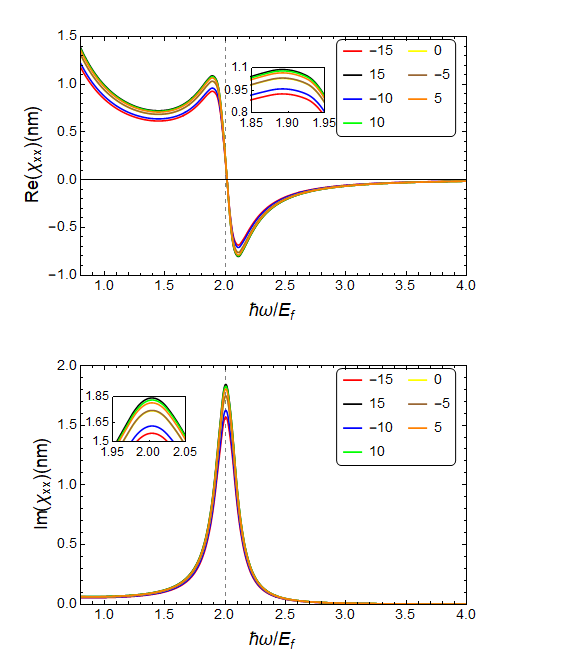}
   \caption{Real and imaginary parts of the linear optical susceptibility component $\chi_{xx}^{(1)}$ as a function of the normalized photon energy $\hbar\omega/E_f$. The labels designate the various values of the strain parameter $w$ (in units of $\text{meV}$). The inset tracks the strain dependence of the peak intensity at the resonance frequency.} 
   \label{2}
\end{figure}

\begin{figure}[t!]
   \centering   
   \includegraphics[width=0.48\textwidth]{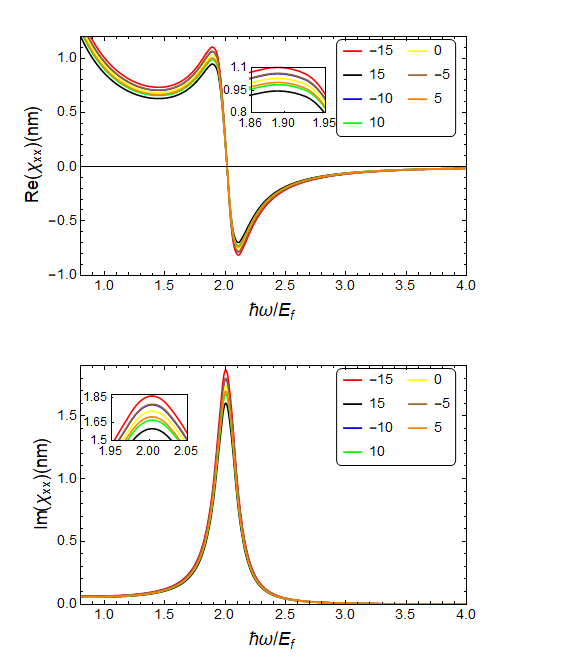}
   \caption{Real and imaginary parts of the linear optical susceptibility component $\chi_{yy}^{(1)}$ along the armchair direction as a function of the normalized photon energy $\hbar\omega/E_f$. The different curves correspond to the indicated values of the strain parameter $w$ (in units of $\text{meV}$).} 
   \label{3}
\end{figure}

The linear optical response of uniaxially strained bilayer graphene is evaluated as a function of the strain parameter $w$ using Eq.~\eqref{pomega}, with the corresponding real and imaginary components presented in Figs.~\ref{2} and \ref{3}. Driven by the microscopic channels defined in Eq.~\eqref{rho1}, the linear susceptibility tensor exhibits a pronounced resonant feature centered at $\hbar\omega/E_f \simeq 2$. This peak marks the onset of direct interband transitions across the Fermi edge. 

For the unbiased bilayer system, the interband excitation energy scales as $E_{cv}(\mathbf{k}) = 2E(\mathbf{k})$, which simplifies to $E_{cv} \approx 2E_f$ at the Fermi surface. Consequently, the threshold condition $\hbar\omega = E_{cv}$ produces a resonant peak at $\hbar\omega \approx 2E_f$, confirming the underlying analytics. The imaginary part of the susceptibility, $\text{Im}[\chi^{(1)}]$, displays a standard Lorentzian absorption profile at this threshold. Concurrently, the real part, $\text{Re}[\chi^{(1)}]$, exhibits a characteristic dispersive line shape that undergoes a sign inversion across the resonance frequency, in strict accordance with the Kramers-Kronig relations.

The strain parameter $w$ primarily modifies the underlying momentum-dependent optical matrix elements, thereby modulating the overall amplitude of the optical response, whereas the resonance frequency remains fundamentally pinned by the Fermi energy $E_f$. The inset of Fig.~\ref{2} illustrates the dependence of the linear optical response along the zigzag direction on the strain parameter at the resonance peak. Specifically, as $w$ increases monotonically from $-15\,\text{meV}$ to $+15\,\text{meV}$, the peak magnitude of the linear response decreases by approximately $23\%$. This trend demonstrates that smaller or more negative values of $w$ yield a higher optical current intensity along this crystallographic direction.  

Conversely, for the linear optical response evaluated along the armchair direction ($y$ axis), the parameter configuration $w = -15\,\text{meV}$ yields the maximum peak amplitude, whereas $w = +15\,\text{meV}$ corresponds to the minimum response (see Fig.~\ref{3}). This behavior indicates that when the strain-shifted Dirac points migrate along the vertical $k_y$ axis, the linear current along the real-space armchair edge is enhanced compared to the configuration where the Dirac points are displaced along the horizontal $k_x$ axis. Furthermore, within the negative strain regime ($w < 0$), an increased separation between the Dirac points correlates with an enhanced linear current intensity; in contrast, within the positive strain regime ($w > 0$), a wider separation suppresses the response. This strain-induced modulation of the linear susceptibility is most pronounced at the interband resonance threshold, as explicitly tracked by the peak-intensity scaling shown in the insets of Figs.~\ref{2} and \ref{3}. 
\subsection{Second-harmonic generation}
\begin{figure}[t!]
   \centering   
   \includegraphics[width=0.48\textwidth]{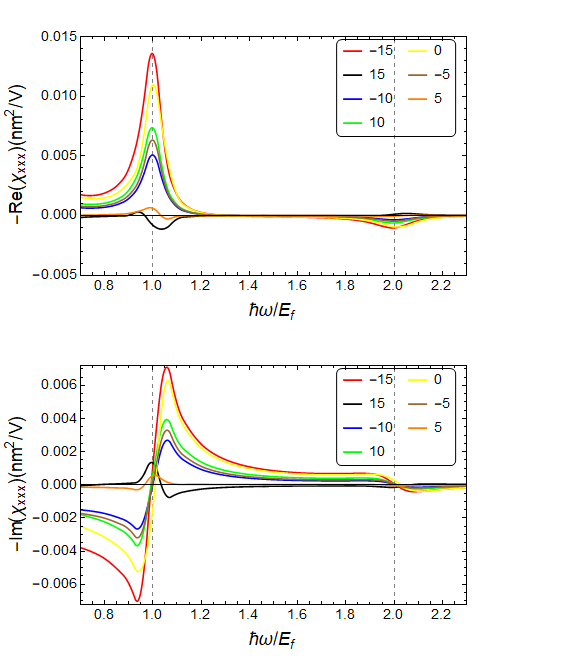}
   \caption{Real and imaginary parts of the second-order nonlinear susceptibility tensor component $\chi_{xxx}^{(2)}$ as a function of the normalized photon energy $\hbar\omega/E_f$. The different curves correspond to the indicated values of the strain parameter $w$ (in units of $\text{meV}$).} 
   \label{4}
\end{figure}

\begin{figure}[t!]
   \centering   
   \includegraphics[width=0.48\textwidth]{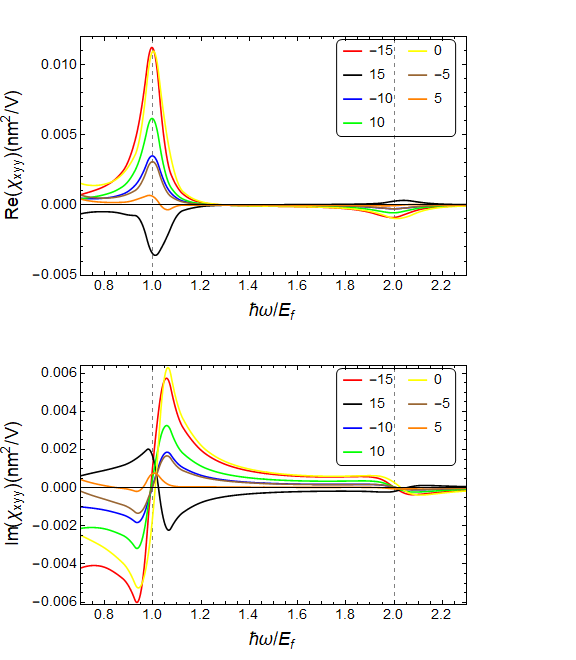}
   \caption{Real and imaginary parts of the transverse second-order susceptibility component $\chi_{xyy}^{(2)}$ as a function of the normalized photon energy $\hbar\omega/E_f$ for various values of the strain parameter $w$ (in units of $\text{meV}$).} 
   \label{5}
\end{figure}

\begin{figure}[t!]
   \centering   
   \includegraphics[width=0.48\textwidth]{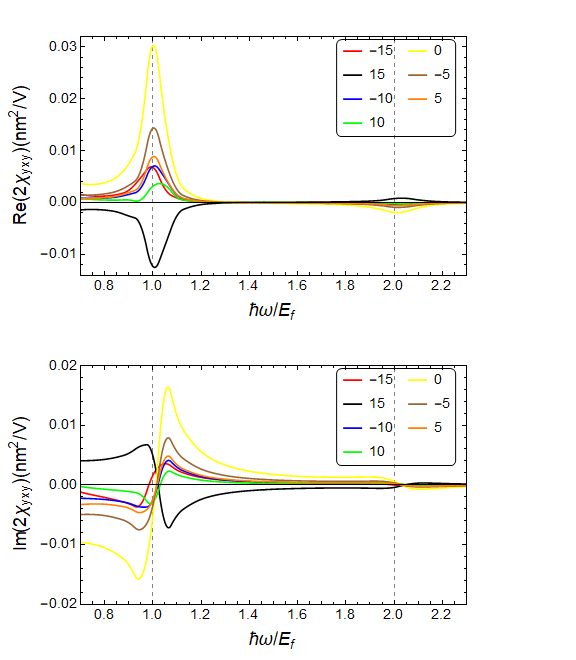}
   \caption{Real and imaginary parts of the second-order nonlinear susceptibility tensor component $2\chi_{yxy}^{(2)}$ as a function of the normalized photon energy $\hbar\omega/E_f$. The individual curves correspond to selected values of the strain parameter $w$ (in units of $\text{meV}$).} 
   \label{6}
\end{figure}

The frequency dependence of the second-order nonlinear susceptibility exhibits a pronounced resonant peak at $\hbar\omega/E_f \simeq 1$ and a secondary, weaker feature at $\hbar\omega/E_f \simeq 2$, as displayed in Figs.~\ref{4} to \ref{6}. These distinct profiles originate directly from the multi-wave mixing resonance denominators embedded within the second-order perturbative density matrix elements [see Eqs.~\eqref{rho2}]. 

For the unbiased bilayer graphene spectrum ($\Delta = 0$) employed in this work, the vertical interband transition energy is given by $E_{cv}(\mathbf{k}) = 2E(\mathbf{k})$. Because the dominant contribution to the nonlinear optical response is restricted to electronic states in the immediate vicinity of the Fermi surface where $E(\mathbf{k}) \approx E_f$, the characteristic interband excitation threshold is established at $2E_f$.

The dominant feature located at $\hbar\omega/E_f \approx 1$ can be directly attributed to the two-photon resonance condition,
\begin{equation}
2\hbar\omega = E_{cv} \approx 2E_f,
\end{equation}
which corresponds to the resonant activation of a two-photon optical transition channel. In contrast, the subdominant peak near $\hbar\omega/E_f \approx 2$ originates from the standard one-photon interband resonance,
\begin{equation}
\hbar\omega = E_{cv} \approx 2E_f.
\end{equation}
The simultaneous emergence of distinct resonant structures at both $\hbar\omega \approx E_f$ and $\hbar\omega \approx 2E_f$ constitutes a hallmark characteristic of second-order nonlinear optical processes. This dual-profile behavior directly reflects the concurrent quantum interference between the fundamental and second-harmonic excitation pathways embedded within the denominators of the nonlinear susceptibility tensor. 

As shown in Figs.~\ref{4} to \ref{6}, the peak magnitude of the induced second-order current densities falls within the same order of magnitude as that predicted for pristine monolayer graphene in Ref.~\cite{tara}. The influence of the phenomenological scattering rate $\gamma$ on the resulting SHG spectrum has been documented previously; specifically, an increase in the damping parameter $\gamma$ systematically suppresses the peak current intensity due to the accelerated dephasing of the coherent electronic state combinations~\cite{tara}.

Although the precise spectral locations of the nonlinear resonances are nearly invariant under variations of the strain parameter $w$, their peak amplitudes are highly sensitive to both the magnitude and sign of $w$. This invariant positioning demonstrates that the applied uniaxial strain primarily modulates the underlying momentum-dependent optical matrix elements, while ruling out significant shifts in the vertical interband transition energies themselves. Microscopically, the strain-induced gauge field $w\sigma_x$ alters the sublattice pseudospin texture of the low-energy Bloch wavefunctions, which fundamentally reshapes the optical transition pathways while concurrently driving the anisotropic distortion of the trigonally warped Fermi surface (see Fig.~\ref{1}). 

Under an in-plane electric field polarized exclusively along the zigzag edge ($x$ axis), the resulting longitudinal second-harmonic current density is governed by the tensor component $\chi_{xxx}^{(2)}$, as plotted in Fig.~\ref{4}. The corresponding orthogonal cross-optical response, characterized by the tensor component $\chi_{xyy}^{(2)}$, is presented in Fig.~\ref{5}. 

A comparative analysis of Figs.~\ref{4} and \ref{5} reveals that the strain configuration $w = -15\,\text{meV}$ yields the maximum nonlinear current amplitude along the $x$ direction. In this regime, the twin Dirac points undergo a relative migration that aligns them vertically along the $k_y$ axis. Conversely, a positive strain parameter ($w = 15\,\text{meV}$) induces a definitive sign inversion in the nonlinear response relative to the other strained configurations, manifesting as a positive peak for $\chi_{xxx}^{(2)}$ and a negative peak for the transverse component $\chi_{xyy}^{(2)}$. Consequently, adjusting the strain field to drive the Dirac points into a horizontal, collinear alignment along the $k_x$ axis directly reverses the sign of the corresponding nonlinear susceptibilities, offering a robust mechanism to switch the directional orientation of the induced second-harmonic current.
  
As illustrated in Fig.~\ref{5}, the maximum transverse nonlinear current is achieved at $w = -15\,\text{meV}$, below which the peak amplitude diminishes monotonically with decreasing values of $|w|$. This spectral enhancement originates from the amplified pseudospin anisotropy and the concurrently enhanced momentum-dependent optical matrix elements along the specific $\mathbf{k}$-space trajectory connecting the strain-shifted Dirac points. Notably, the transverse current undergoes a complete directional reversal for $w = +15\,\text{meV}$, with its peak magnitude at the resonance threshold being suppressed by approximately a factor of two compared to the corresponding $w = -15\,\text{meV}$ configuration. For intermediate values of the strain parameter, the nonlinear optical profiles evolve continuously, demonstrating the highly tunable nature of the transverse harmonic response via deterministic strain engineering. 

When the incident optical field possesses arbitrary in-plane polarization with concurrent nonzero $E_x$ and $E_y$ components, it drives an additional second-harmonic current density along the transverse $y$ direction. This specific non-linear response channel is governed by the cross-polarization susceptibility tensor component $2\chi_{yxy}^{(2)}$, as presented in Fig.~\ref{6}.

As illustrated in Fig.~\ref{6}, the introduction of uniaxial strain generally suppresses the overall magnitude of this transverse susceptibility component compared to the pristine layout. Nevertheless, within the strained regime, the parameter configuration $w = +15\,\text{meV}$ yields the dominant response, manifesting as a pronounced negative peak that indicates a highly directional induced current oriented along the $-y$ axis. 

Within the parameter space explored in this work, the second-order nonlinear profiles do not obey a simple or universal scaling relation with respect to the magnitude of the strain parameter $w$, as evidenced by the complex variations across Figs.~\ref{4} to \ref{6}. 

For the unstrained configuration ($w = 0$), the total macroscopic SHG signal arises strictly from the uncompensated contribution between opposing valleys, where $\chi^{(2)}(\mathbf{K}) \neq -\chi^{(2)}(\mathbf{K}')$, establishing a valley-contrast mechanism analogous to that reported for monolayer graphene~\cite{tara}. Furthermore, a comparative inspection of Figs.~\ref{2} to \ref{6} reveals that the induced second-order nonlinear response remains approximately two orders of magnitude smaller than its linear counterpart across the entire spectral range, consistent with the standard perturbative hierarchy of optical susceptibilities in 2D systems.

The calculated peak nonlinear susceptibility reaches values of order
$10^{-2}\,\mathrm{nm}^2/\mathrm{V}$, which lie within the range predicted
for several graphene-based nonlinear optical platforms~\cite{tara}.
Although this value is smaller than the strongest SHG responses reported
for inversion-symmetry-broken bilayer graphene~\cite{Brun2015}, the
present mechanism offers a distinct advantage: the nonlinear current can
be reversibly switched through the application of uniaxial strain. An experimentally accessible signature of the present mechanism is the
reversal of the SHG current direction upon changing the sign of the
uniaxial strain parameter. Since the resonance frequencies remain
essentially fixed while the nonlinear susceptibility changes sign,
polarization-resolved SHG measurements could directly distinguish this
effect from conventional intensity modulation. In contrast to gate-controlled inversion-symmetry breaking, strain
engineering modifies the nonlinear response through the deformation of
the low-energy band structure and pseudospin texture. As a result,
mechanical control provides an independent tuning parameter that can be
combined with electrostatic and optical approaches for manipulating
valley-dependent nonlinear optical phenomena.
The theoretical framework developed here can be extended to investigate
higher-order harmonic generation, nonlinear Hall effects, and other
strain-engineered valleytronic phenomena in multilayer graphene systems.

\section{Conclusion}
In summary, we have theoretically investigated the linear and second-harmonic optical responses of valley-polarized bilayer graphene subjected to uniaxial strain within a low-energy quantum kinetic framework. Our findings demonstrate that the complex interplay among intrinsic trigonal warping, valley polarization, and strain-induced electronic anisotropy gives rise to a finite, highly tunable optoelectronic profile. Notably, the SHG spectrum exhibits a significantly heightened sensitivity to variations in the strain parameter compared to the linear optical response, providing an effective lever to completely reverse the directional orientation of the induced nonlinear current.

Microscopically, the strain parameter $w$ breaks the structural symmetry by distorting and shifting the low-energy Dirac cones, which fundamentally reshapes the sublattice pseudospin texture and alters the underlying momentum-dependent interband transition pathways. Depending on the magnitude and sign of $w$, the resulting SHG signal can be deterministically enhanced or suppressed. Within the parameter regime explored here, the nonlinear susceptibility profiles evolve continuously without adhering to a simple or universal scaling relation with $w$, highlighting the intricate quantum interference between competing transition channels. 

Furthermore, while the linear response exhibits a characteristic resonance threshold at the standard interband edge $\hbar\omega \sim 2E_f$, the second-order nonlinear response features a dual-peaked structure in the mid-infrared regime with pronounced resonant enhancements centered near both $\hbar\omega \sim E_f$ and $\hbar\omega \sim 2E_f$, whose spectral positions remain largely invariant under strain.  Taken together, these results demonstrate that strain engineering provides an effective route for controlling valley-dependent nonlinear optical responses in bilayer graphene.

\bibliographystyle{apsrev4-2}
\bibliography{bib}

\end{document}